# Achievable Rates for a Two-Relay Network with Relays-Transmitter Feedbacks

Mohammad Ali Tebbi, *Student Member, IEEE*
Mahmoud Ahmadian Attari
Coding and Cryptography Lab. (CCL), Department of ECE
K.N. Toosi University of Technology, Tehran, Iran
m.a.tebbi@ee.kntu.ac.ir, mahmoud@eetd.kntu.ac.ir

Mahtab Mirmohseni, Mohammad Reza Aref
Information Systems and Security Lab. (ISSL)
Department of EE, Sharif University of Techology
Tehran, Iran
mirmohseni@ee.sharif.edu, aref@sharif.edu

*Abstract*—We consider a relay network with two relays and two feedback links from the relays to the sender. To obtain the achievability results, we use the compress-and-forward and the decode-and-forward strategies to superimpose facility and cooperation analogue to what proposed by Cover and El Gamal for a relay channel. In addition to random binning, we use deterministic binning to perform restricted decoding. We show how to use the feedback links for cooperation between the sender and the relays to transmit the information which is compressed in the sender and the relays.

## I. INTRODUCTION

The relay channel with feedback was first considered by Cover and El Gamal in [1]. In their channel model, there were feedback links from the receiver to both the sender and the relay and from the relay to the sender, referred to as complete feedback [2]. It was shown that the presence of complete feedback or partial feedback from the receiver to the relay makes the relay channel a physically degraded relay channel, thus the cut-set upper bound would be achievable [1]. The relay channel with a partial feedback from the receiver or the relay to the sender has been investigated in [3] and [4]. It has been shown that neither the receiver-transmitter nor the relay-transmitter feedback can improve the capacity of the physically degraded and the semi-deterministic relay channels [4].

The relay network was first introduced in [5], where the capacity of a general relay network with complete feedback, i.e. feedbacks from the receiver to all relays and the sender, and from each relay to the sender and the previous relays, has been derived and shown that the complete feedback can increase the capacity. The presence of Partial feedbacks from the receiver to the relays and from each relay to the previous ones make the relay network a physically degraded relay network, thus cannot increase the capacity [6].

In [7] and [8], relay networks with parallel relaying have been considered. In parallel relaying, there is no straight link between the sender and the receiver. Also, the relays do not interchange any information. In [9], some cooperative strategies for relay networks have been discussed and reviewed. Additionally, the authors in [9] generalize the compress-and-forward strategy to the relay networks. Symmetric relay network has been introduced in [10]. In a symmetric two-relay network, there is a straight link from the sender to the receiver and each relay can completely decode the message transmitted by the other relay in addition to the part of the message transmitted by the sender.

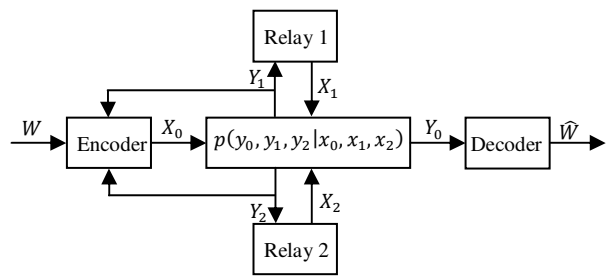

Fig. 1. A Two-Relay Network with Relays-Transmitter Feedbacks.

In [11], the authors have derived an achievable rate for a two-relay network with *receiver-transmitter* feedback. The achievability result in [11] have been based on compress-and-forward relaying scheme. It has been showed that in [11], how to use the feedback information to build a two-level cooperation between the sender and the relays. However, no achievable rate expression has yet been obtained for relay networks with partial feedback from the relays to the sender.

In this paper, we consider a relay network with two relays and partial feedbacks from the relays to the sender. In our proposed model depicted in Fig. 1, there is a feed-forward link from the sender to the receiver and two feedback links from the relays to the sender. Both of the relays help the receiver to solve his uncertainty about the sender. Each relay tries to send the information about the sender's messages to the receiver as much as possible through the direct link between the relay and the receiver. However, the relays have no information interchange. since the sender knows the information received by the relays, The feedback links develop a complete cooperation between the sender and the relays. Thus, they cooperate in transmitting the compressed information to the receiver. For this model, we present two achievable rates. Our first result is based on the compress-and-forward coding scheme [1] and random partitioning [12]. The second result is based on the compress-and-forward coding scheme [1] combined with the decode-and-forward coding scheme. To perform restricted decoding [13], we use the deterministic partitioning in addition to the random partitioning.

The rest of the paper is organized as fallows. Section II, introduces the network model and definitions. In Section III, we present two achievable rates obtained for the model. In Section IV, the achievability of the rates reported in Section III are proved. Finally, in Section V, we conclude the paper.

## II. PRELIMINARIES AND DEFINITIONS

In this paper, upper case letters (e.g., $X$) are used to denote Random Variables (RVs) while their realizations are denoted by lower case letters (e.g., $x$). The alphabet of a random variable $X$ will be designated by a calligraphic letter $\mathcal{X}$. $x_i^{j(k)}$ indicates the sequence of $\left(x_{i,1}^{(k)}, x_{i,2}^{(k)}, \ldots, x_{i,j}^{(k)}\right)$, where $k$ denotes the block number of transmission. $p_X(x)$ denotes the probability mass function (p.m.f) of $X$ on a set $\mathcal{X}$, where occasionally subscript $X$ is omitted.

*Definition 1:* The discrete memoryless two-relay network $(\mathcal{X}_0 \times \mathcal{X}_1 \times \mathcal{X}_2, p(y_0, y_1, y_2 | x_0, x_1, x_2), \mathcal{Y}_0 \times \mathcal{Y}_1 \times \mathcal{Y}_2)$ with relays-transmitter feedbacks depicted in Fig. 1, consists of a sender $X_0 \in \mathcal{X}_0$, a receiver $Y_0 \in \mathcal{Y}_0$, relay senders $X_1 \in \mathcal{X}_1$ and $X_2 \in \mathcal{X}_2$, relay receivers $Y_1 \in \mathcal{Y}_1$ and $Y_2 \in \mathcal{Y}_2$, and a family of conditional probability mass functions $p(y_0, y_1, y_2 | x_0, x_1, x_2)$ on $\mathcal{Y}_0 \times \mathcal{Y}_1 \times \mathcal{Y}_2$ one for each $(x_0, x_1, x_2) \in \mathcal{X}_0 \times \mathcal{X}_1 \times \mathcal{X}_2$. An $(M, n)$ code, for this network consists of a message set $\mathcal{M} = \{1, 2, \ldots, M\}$, an encoding function $x_0 : \mathcal{M} \times \mathcal{Y}_1^{t-1} \times \mathcal{Y}_2^{t-1} \to \mathcal{X}_0$ for $t = 1, \ldots, n$, a set of relay functions $\{f_{ij}\}$ such that $x_{ij} = f_{ij}(y_{i1}, y_{i2}, \ldots, y_{i,j-1})$ for $1 \leq j \leq n$, $i = \{1,2\}$, and a decoding function $g : \mathcal{Y}_0 \to \mathcal{M}$. A rate $R = \frac{1}{n} \log M$ is achievable if there is an $(M, n)$ code with $M \geq 2^{nR}$ such that $\bar{p}_e^n = \Pr\{g(Y_0^n) \neq W | W = w, w \in \mathcal{M}\} < \varepsilon$, for any $\varepsilon > 0$ and for sufficiently large $n$.

## III. MAIN RESULTS

In this Section, we present two achievable rates concerning a relay network with two relays and feedback links from both of the relays to the sender as depicted in Fig. 1.

*Theorem 1:* Consider the discrete memoryless relay network with two relays $(\mathcal{X}_0 \times \mathcal{X}_1 \times \mathcal{X}_2, p(y_0, y_1, y_2 | x_0, x_1, x_2), \mathcal{Y}_0 \times \mathcal{Y}_1 \times \mathcal{Y}_2)$ and causal noiseless feedbacks from both of the relays to the sender defined in Section II. Then, the rate $\bar{\bar{R}}$ defined by

$$\bar{\bar{R}} = \sup_{p(x_1, x_2, x_0, y_0, y_1, y_2, \hat{y}_1, \hat{y}_2)} I(X_0; Y_0, \hat{Y}_1, \hat{Y}_2 | X_1, X_2) + I(X_1; X_2)$$

(1)

is achievable subject to the constraints

$$I(\hat{Y}_1; Y_1 | X_1) + I(\hat{Y}_1; Y_2, X_2 | X_1, Y_1) < I(X_1; Y_0, X_2) + I(\hat{Y}_1; Y_0 | X_1)$$

(2)

$$I(\hat{Y}_2; Y_2 | X_2) + I(\hat{Y}_2; Y_1, X_1 | X_2, Y_2) < I(X_2; Y_0, X_1) + I(\hat{Y}_2; Y_0 | X_2)$$

(3)

$$I(\hat{Y}_1; Y_1 | X_1) + I(\hat{Y}_2; Y_2 | X_2) + I(\hat{Y}_1; Y_2, X_2 | X_1, Y_1) + I(\hat{Y}_2; Y_1, X_1 | X_2, Y_2)$$
$$< \min \begin{Bmatrix} I(X_1, X_2; Y_0) + I(\hat{Y}_1; Y_0 | X_1) + I(\hat{Y}_2; Y_0 | X_2), \\ I(X_1; Y_0, X_2) + I(X_2; Y_0, X_1) + I(\hat{Y}_1; Y_0 | X_1) + I(\hat{Y}_2; Y_0 | X_2) \end{Bmatrix}$$

(4)

where the supremum is taken over all joint p.m.fs on $\mathcal{X}_0 \times \mathcal{X}_1 \times \mathcal{X}_2 \times \mathcal{Y}_1 \times \mathcal{Y}_2 \times \hat{\mathcal{Y}}_1 \times \hat{\mathcal{Y}}_2 \times \mathcal{Y}_0$ of the form

$$p(x_0, x_1, x_2, y_1, y_2, \hat{y}_1, \hat{y}_2, y_0)$$
$$= p(x_1)p(x_2)p(x_0|x_1, x_2)p(y_0, y_1, y_2|x_0, x_1, x_2)$$
$$\cdot p(\hat{y}_1|y_1, x_1)p(\hat{y}_2|y_2, x_2)$$

*Remark 1:* Constraints (2), (3), and (4) are dictated by the decoding procedure in the receiver and reflect the minimal compression ratio sustainable by the receiver, taking into account the help it gets from the sender and the first relay, the sender and the second relay, and the sender and both of the relays, respectively.

*Theorem 2:* Consider the discrete memoryless relay network $(\mathcal{X}_0 \times \mathcal{X}_1 \times \mathcal{X}_2, p(y_0, y_1, y_2 | x_0, x_1, x_2), \mathcal{Y}_0 \times \mathcal{Y}_1 \times \mathcal{Y}_2)$ with two relays and causal noiseless feedbacks from both of the relays to the sender defined in Section II. Then, the rate $\bar{R}$ defined by

$$\bar{R} = \sup_{p(x_1, x_2, v_1, v_2, x_0, y_0, y_1, y_2, \hat{y}_1, \hat{y}_2)}$$

$$I(X_0; Y_0, \hat{Y}_1, \hat{Y}_2 | X_1, X_2, V_1, V_2) + I(X_1, V_1; X_2, V_2) + R_{21} + R_{22}$$

(5)

is achievable subject to the constraints

$$R_{21} < \min \begin{Bmatrix} I(V_1; Y_1 | X_1), \\ I(V_1; Y_0 | X_1) + I(X_1; Y_0, X_2) + I(\hat{Y}_1; Y_0 | X_1, V_1) \\ -I(\hat{Y}_1; Y_1, Y_2, X_2, V_2 | X_1, V_1) \end{Bmatrix}$$

(6)

$$R_{22} < \min \begin{Bmatrix} I(V_2; Y_2 | X_2), \\ I(V_2; Y_0 | X_2) + I(X_2; Y_0, X_1) + I(\hat{Y}_2; Y_0 | X_2, V_2) \\ -I(\hat{Y}_2; Y_1, Y_2, X_1, V_1 | X_2, V_2) \end{Bmatrix}$$

(7)

$$R_{21} + R_{22}$$
$$< \min \begin{Bmatrix} I(V_1; Y_0 | X_1) + I(V_2; Y_0 | X_2) + I(X_1, X_2; Y_0) \\ +I(\hat{Y}_1; Y_0 | X_1, V_1) + I(\hat{Y}_2; Y_0 | X_2, V_2) \\ -I(\hat{Y}_1; Y_1, Y_2, X_2, V_2 | X_1, V_1) - I(\hat{Y}_2; Y_1, Y_2, X_1, V_1 | X_2, V_2), \\ I(V_1; Y_0 | X_1) + I(V_2; Y_0 | X_2) + I(X_1; Y_0, X_2) \\ +I(X_2; Y_0, X_1) + I(\hat{Y}_1; Y_0 | X_1, V_1) + I(\hat{Y}_2; Y_0 | X_2, V_2) \\ -I(\hat{Y}_1; Y_1, Y_2, X_2, V_2 | X_1, V_1) - I(\hat{Y}_2; \hat{Y}_1, Y_1, Y_2, X_1, V_1 | X_2, V_2) \end{Bmatrix}$$

(8)

where the supremum is taken over all joint p.m.fs on $\mathcal{X}_0 \times \mathcal{X}_1 \times \mathcal{X}_2 \times \mathcal{V}_1 \times \mathcal{V}_2 \times \mathcal{Y}_1 \times \mathcal{Y}_2 \times \hat{\mathcal{Y}}_1 \times \hat{\mathcal{Y}}_2 \times \mathcal{Y}_0$ of the form

$$p(x_1, x_2, v_1, v_2, x_0, y_0, y_1, y_2, \hat{y}_1, \hat{y}_2)$$
$$= p(x_1)p(x_2)p(v_1|x_1)p(v_2|x_2)p(x_0|x_1, x_2, v_1, v_2)$$
$$\cdot p(y_0, y_1, y_2|x_0, x_1, x_2)p(\hat{y}_1|x_1, v_1, y_1)p(\hat{y}_2|x_2, v_2, y_2)$$

*Remark 2:* The first terms in the constraints (6) and (7) are dictated by the decoding procedure in the first and the second relays, respectively, due to the decode-and-forward coding scheme. The second terms in the constraints (6) and (7), and the constraint (8) are dictated by the decoding procedure in the receiver and reflect the minimal compression ratio sustainable by the receiver, taking into account the help it gets from the sender and the first relay, the sender and the second relay, and the sender and both of the relays, respectively.

*Remark 3:* To compare the two achievable rates, let $V_1 = X_1$, $V_2 = X_2$, $R_{21} = 0$, $R_{22} = 0$, $p(\hat{y}_1|x_1, v_1, y_1) = p(\hat{y}_1|x_1, y_1)$, and $p(\hat{y}_2|x_2, v_2, y_2) = p(\hat{y}_2|x_2, y_2)$ in the achievable rate region of theorem 2. In which case (6), (7), and (8) become the same as (2), (3), and (4), respectively. The achievable rate $\bar{R}$ becomes the same as $\bar{\bar{R}}$, where the supremum is taken over all laws

$$p(x_0, x_1, x_2, y_1, y_2, \hat{y}_1, \hat{y}_2, y_0)$$
$$= p(x_1)p(x_2)p(x_0|x_1, x_2)p(y_0, y_1, y_2|x_0, x_1, x_2)$$
$$\cdot p(\hat{y}_1|y_1, x_1)p(\hat{y}_2|y_2, x_2)$$

which is precisely the definition of $\bar{\bar{R}}$. Therefore, $\bar{R}$ includes $\bar{\bar{R}}$.

## IV. PROOFS

To prove the achievability of rate $\bar{\bar{R}}$ in Theorem 1, we use the compress-and-forward coding scheme [1] based on block Markov superposition encoding and random binning proof of the source coding theorem of Slepian-Wolf [12]. Achievability of rate $\bar{R}$ is proved with combining the compress-and-forward and the decode-and-forward coding schemes. In addition to random binning, we utilize nonrandom binning to decrease the complexity of list coding techniques. With this, we can perform the restricted decoding, instead of list decoding and lexicographical indexing. This method was introduced in [13] for multiple-access channel with partial feedback. In some of the decoding steps, we use joint decoding technique [14].

*Proof of Theorem 1:* Consider a block Markov encoding scheme where a sequence of $B - 1$ messages $w^{(b)} \in [1, 2^{n\bar{R}}]$ for $b = 1, 2, \ldots, B - 1$ is transmitted in $B$ blocks, each of $n$ symbols. As $B \to \infty$, the rate $\bar{\bar{R}}(B-1)/B$ is arbitrarily close to $\bar{\bar{R}}$.

***Random coding:*** Generate $2^{nR_{s_1}}$ i.i.d sequences $x_1^n$, each with probability $p(x_1^n) = \prod_{i=1}^n p(x_{1i})$ and label them as $x_1^n(s_1)$, where $s_1 \in [1, 2^{nR_{s_1}}]$. Generate $2^{nR_{s_2}}$ i.i.d sequences $x_2^n$, each with probability $p(x_2^n) = \prod_{i=1}^n p(x_{2i})$ and label them as $x_2^n(s_2)$, where $s_2 \in [1, 2^{nR_{s_2}}]$. For each $(x_1^n, x_2^n)$, generate $2^{n\bar{R}}$ i.i.d sequences $x_0^n$, each with probability $p(x_0^n|x_1^n, x_2^n) = \prod_{i=1}^n p(x_{0i}|x_{1i}, x_{2i})$ and label them as $x_0^n(w, s_1, s_2)$, where $w \in [1, 2^{n\bar{R}}]$. For each $x_1^n$, choose $2^{n\hat{R}_1}$ i.i.d sequences $\hat{y}_1^n$, each with probability $p(\hat{y}_1^n|x_1^n) = \prod_{i=1}^n p(\hat{y}_{1i}|x_{1i})$, where for $x_1 \in \mathcal{X}_1$ and $\hat{y}_1 \in \hat{\mathcal{Y}}_1$ we define

$$p(\hat{y}_1^n|x_1^n)$$
$$= \sum_{x_0, y_0, y_1, y_2} p(x_0|x_1, x_2)p(y_0, y_1, y_2|x_0, x_1, x_2)p(\hat{y}_1|x_1, y_1)$$

and label them as $\hat{y}_1^n(z_1|s_1)$, where $z_1 \in [1, 2^{n\hat{R}_1}]$. For each $x_2^n$, choose $2^{n\hat{R}_2}$ i.i.d sequences $\hat{y}_2^n$, each with probability $p(\hat{y}_2^n|x_2^n) = \prod_{i=1}^n p(\hat{y}_{2i}|x_{2i})$, where for $x_2 \in \mathcal{X}_2$ and $\hat{y}_2 \in \hat{\mathcal{Y}}_2$ we define

$$p(\hat{y}_2^n|x_2^n)$$
$$= \sum_{x_0, y_0, y_1, y_2} p(x_0|x_1, x_2)p(y_0, y_1, y_2|x_0, x_1, x_2)p(\hat{y}_2|x_2, y_2)$$

and label them as $\hat{y}_2^n(z_2|s_2)$, where $z_2 \in [1, 2^{n\hat{R}_2}]$.

***Partitioning:***
1. Randomly partition the set $\{1, 2, \ldots, 2^{n\hat{R}_1}\}$ into $2^{nR_{s_1}}$ cells $\mathcal{S}_{s_1}$ for $s_1 \in \{1, 2, \ldots, 2^{nR_{s_1}}\}$, and the set $\{1, 2, \ldots, 2^{n\hat{R}_2}\}$ into $2^{nR_{s_2}}$ cells $\mathcal{S}_{s_2}$ for $s_2 \in \{1, 2, \ldots, 2^{nR_{s_2}}\}$.

***Encoding:***
Let $w^{(b)}$ be the new message to be sent in block $b$. Assume that $(\hat{y}_1^n(z_1^{(b-1)}|s_1^{(b-1)}), y_1^{n(b-1)}, x_1^{n(b-1)})$ are jointly $\epsilon$-typical, and $(\hat{y}_2^n(z_2^{(b-1)}|s_2^{(b-1)}), y_2^{n(b-1)}, x_2^{n(b-1)})$ are jointly $\epsilon$-typical. Then, the codewords transmitted by the first and the second relays in block $b$ are

$$x_1^n(s_1^{(b)}) = x_1^n(s_1(z_1^{(b-1)}))$$

and

$$x_2^n(s_2^{(b)}) = x_2^n(s_2(z_2^{(b-1)})),$$

respectively, while the codeword transmitted by the sender is

$$x_0^n(w^{(b)}, s_1^{(b)}, s_2^{(b)}) = x_0^n(w^{(b)}, s_1(z_1^{(b-1)}), s_2(z_2^{(b-1)})).$$

***Decoding:***
*At the first relay:* At the end of block $b$, $b = 1, 2, \ldots, B - 1$, the first relay knowing $s_1^{(b)}$ and upon receiving $y_1^{n(b)}$, decides that $z_1^{(b)}$ is received if $(\hat{y}_1^n(z_1^{(b)}|s_1^{(b)}), y_1^{n(b)}, x_1^n(s_1^{(b)}))$ are jointly $\epsilon$-typical. There exists such a $z_1^{(b)}$ with high probability if

$$\hat{R}_1 > I(\hat{Y}_1; Y_1|X_1) \qquad (9)$$

and $n$ is sufficiently large.

*At the second relay:* At the end of block $b$, $b = 1, 2, \ldots, B - 1$, the second relay knowing $s_2^{(b)}$ and upon receiving $y_2^{n(b)}$, decides that $z_2^{(b)}$ is received if $(\hat{y}_2^n(z_2^{(b)}|s_2^{(b)}), y_2^{n(b)}, x_2^n(s_2^{(b)}))$ are jointly $\epsilon$-typical. There exists such a $z_2^{(b)}$ with high probability if

$$\hat{R}_2 > I(\hat{Y}_2; Y_2|X_2) \qquad (10)$$

and $n$ is sufficiently large.
Using covering lemma [14], (9) and (10) are proved.

*At the sender:* At the end of block $b$, $b = 1, 2, \ldots, B - 1$, the sender knowing $(s_1^{(b)}, s_2^{(b)})$ and upon receiving $y_1^{n(b)}$ and $y_2^{n(b)}$, decides that the pair $(z_1^{(b)}, z_2^{(b)})$ is received if

$\left(\hat{y}_1^n(z_1^{(b)}|s_1^{(b)}), \hat{y}_2^n(z_2^{(b)}|s_2^{(b)}), y_1^{n(b)}, y_2^{n(b)}, x_1^n(s_1^{(b)}), x_2^n(s_2^{(b)})\right)$ are jointly $\epsilon$-typical. There exists such a pair $(z_1^{(b)}, z_2^{(b)})$ with high probability if

$$\hat{R}_1 > I(\hat{Y}_1; Y_1, Y_2, X_2|X_1) \quad (11)$$
$$\hat{R}_2 > I(\hat{Y}_2; Y_1, Y_2, X_1|X_2) \quad (12)$$
$$\hat{R}_1 + \hat{R}_2 > I(\hat{Y}_1; Y_1, Y_2, X_2|X_1) + I(\hat{Y}_2; \hat{Y}_1, Y_1, Y_2, X_1|X_2) \quad (13)$$

and $n$ is sufficiently large.
Equations (11)-(13) are proved using multivariate covering lemma [14].

*Remark 4:* By the decoding steps explained above, the sender, the first, and the second relays cooperate with respectively sending $\left(s_1^{(b+1)}, s_2^{(b+1)}\right) = \left(s_1(z_1^{(b)}), s_2(z_2^{(b)})\right)$, $s_1^{(b+1)} = s_1(z_1^{(b)})$ and $s_2^{(b+1)} = s_2(z_2^{(b)})$ during block $b+1$.

*At the receiver:* At the end of block $b$, the receiver looks for a unique pair $(\hat{s}_1^{(b)}, \hat{s}_2^{(b)})$, such that $\left(x_1^n(\hat{s}_1^{(b)}), x_2^n(\hat{s}_2^{(b)}), y_0^{n(b)}\right)$ are jointly $\epsilon$-typical. Using packing lemma [14] and joint decoding [14, 11], for sufficiently large $n$, the decoding error in this step is arbitrarily small if

$$R_{s_1} < I(X_1; Y_0, X_2) \quad (14)$$
$$R_{s_2} < I(X_2; Y_0, X_1) \quad (15)$$
$$R_{s_1} + R_{s_2} < I(X_1, X_2; Y_0) \quad (16)$$

Then, the receiver considers block $b-1$ and calculates his ambiguity sets $\mathcal{L}_{1D}(y_0^{n(b-1)})$ and $\mathcal{L}_{2D}(y_0^{n(b-1)})$ of $\hat{z}_{1D}^{(b-1)}$ and $\hat{z}_{2D}^{(b-1)}$ such that $\hat{z}_{1D}^{(b-1)} \in \mathcal{L}_{1D}(y_0^{n(b-1)})$ and $\hat{z}_{2D}^{(b-1)} \in \mathcal{L}_{2D}(y_0^{n(b-1)})$ if $\left(x_1^n(\hat{s}_1^{(b-1)}), y_0^{n(b-1)}, \hat{y}_1^n(\hat{z}_{1D}^{(b-1)}|\hat{s}_1^{(b-1)})\right)$ are jointly $\epsilon$-typical, and $\left(x_2^n(\hat{s}_2^{(b-1)}), y_0^{n(b-1)}, \hat{y}_2^n(\hat{z}_{2D}^{(b-1)}|\hat{s}_2^{(b-1)})\right)$ are jointly $\epsilon$-typical. The receiver declares that $\hat{z}_{1D}^{(b-1)}$ and $\hat{z}_{2D}^{(b-1)}$ were sent in block $b-1$ iff there are unique

$$\hat{z}_{1D}^{(b-1)} \in \mathcal{S}_{\hat{s}_1^{(b)}} \cap \mathcal{L}_{1D}(y_0^{n(b-1)})$$

and

$$\hat{z}_{2D}^{(b-1)} \in \mathcal{S}_{\hat{s}_2^{(b)}} \cap \mathcal{L}_{2D}(y_0^{n(b-1)}),$$

respectively. For sufficiently large $n$, the decoding error in this step is arbitrarily small if

$$\hat{R}_1 < I(\hat{Y}_1; Y_0|X_1) + R_{s_1} \quad (17)$$
$$\hat{R}_2 < I(\hat{Y}_2; Y_0|X_2) + R_{s_2} \quad (18)$$

The proofs of (17) and (18) are similar to those in [1] for the compressed information of the relay. For a similar proof, refer to [11].

Then, the receiver declares that $\hat{w}^{(b-1)}$ was sent in block $b-1$ if $\left(x_1^n(\hat{s}_1^{(b-1)}), x_2^n(\hat{s}_2^{(b-1)}), x_0^n(\hat{w}^{(b-1)}, \hat{s}_1^{(b-1)}, \hat{s}_2^{(b-1)}), y_0^{n(b-1)}, \hat{y}_1^n(\hat{z}_{1D}^{(b-1)}|\hat{s}_1^{(b-1)}), \hat{y}_2^n(\hat{z}_{2D}^{(b-1)}|\hat{s}_2^{(b-1)})\right)$ are jointly $\epsilon$-typical. Using packing lemma [14], for sufficiently large $n$, the decoding error in this step is arbitrarily small if

$$\bar{\bar{R}} < I(X_0; Y_0, \hat{Y}_1, \hat{Y}_2|X_1, X_2) + I(X_1; X_2) \quad (19)$$

Combining (9)-(18) and applying Fourier-Motzkin elimination or the algorithm introduced in [15] and eliminating redundant inequalities, constraints (2)-(4) are derived.

*Proof of Theorem 2:* Consider a block Markov encoding scheme where a sequence of $B$ messages $w^{(b)} = \left(w_1^{(b)}, w_{21}^{(b)}, w_{22}^{(b)}\right)$ such that $w^{(b)} \in [1, 2^{n\bar{R}}]$, $w_1^{(b)} \in [1, 2^{nR}]$, $w_{21}^{(b)} \in [1, 2^{nR_{21}}]$, and $w_{22}^{(b)} \in [1, 2^{nR_{22}}]$ for $b = 1,2,\dots,B$ is transmitted in $B + 1$ blocks, each of $n$ symbols. Setting $\bar{R} = R_1 + R_{21} + R_{22}$, as $B \to \infty$, the rate $\bar{R}B/(B+1)$ is arbitrarily close to $\bar{R}$.

***Random coding:*** Generate $2^{n(R_{011}+R_{012})}$ i.i.d sequences $x_1^n$, each with probability $p(x_1^n) = \prod_{i=1}^n p(x_{1i})$ and label them as $x_1^n(w_{01})$, where $w_{01} = (w_{011}, w_{012})$, $w_{011} \in [1, 2^{nR_{011}}]$, and $w_{012} \in [1, 2^{nR_{012}}]$. Generate $2^{n(R_{021}+R_{022})}$ i.i.d sequences $x_2^n$, each with probability $p(x_2^n) = \prod_{i=1}^n p(x_{2i})$ and label them as $x_2^n(w_{02})$, where $w_{02} = (w_{021}, w_{022})$, $w_{021} \in [1, 2^{nR_{021}}]$, and $w_{022} \in [1, 2^{nR_{022}}]$. For each $x_1^n$, generate $2^{nR_{21}}$ i.i.d sequences $v_1^n$, each with probability $p(v_1^n|x_1^n) = \prod_{i=1}^n p(v_{1i}|x_{1i})$ and label them as $v_1^n(w_{21}, w_{01})$, where $w_{21} \in [1, 2^{nR_{21}}]$. For each $x_2^n$, generate $2^{nR_{22}}$ i.i.d sequences $v_2^n$, each with probability $p(x_2^n|v_2^n) = \prod_{i=1}^n p(x_{2i}|v_{2i})$ and label them as $v_2^n(w_{22}, w_{02})$, where $w_{22} \in [1, 2^{nR_{22}}]$. For each $(x_1^n, v_1^n, x_2^n, v_2^n)$, generate $2^{nR_1}$ i.i.d sequences $x_0^n$, each with probability $p(x_0^n|x_1^n, v_1^n, x_2^n, v_2^n) = \prod_{i=1}^n p(x_{0i}|x_{1i}, v_{1i}, x_{2i}, v_{2i})$ and label them as $x_0^n(w_1, w_{21}, w_{22}, w_{01}, w_{02})$, where $w_1 \in [1, 2^{nR_1}]$. For each $(x_1^n, v_1^n)$, choose $2^{n\hat{R}_1}$ i.i.d sequences $\hat{y}_1^n$, each with probability $p(\hat{y}_1^n|x_1^n, v_1^n) = \prod_{i=1}^n p(\hat{y}_{1i}|x_{1i}, v_{1i})$, where for $x_1 \in \mathcal{X}_1$, $\hat{y}_1 \in \hat{\mathcal{Y}}_1$ and $v_1 \in \mathcal{V}_1$ we define

$$p(\hat{y}_1^n|x_1^n, v_1^n) = \sum_{x_0, y_0, y_1, y_2} p(x_0|x_1, x_2, v_1, v_2) p(y_0, y_1, y_2|x_0, x_1, x_2) p(\hat{y}_1|x_1, v_1, y_1)$$

and label them as $\hat{y}_1^n(z_1|w_{21}, w_{01})$, where $z_1 \in [1, 2^{n\hat{R}_1}]$. For each $(x_2^n, v_2^n)$, choose $2^{n\hat{R}_2}$ i.i.d sequences $\hat{y}_2^n$, each with probability $p(\hat{y}_2^n|x_2^n, v_2^n) = \prod_{i=1}^n p(\hat{y}_{2i}|x_{2i}, v_{2i})$ where, for $x_2 \in \mathcal{X}_2$, $\hat{y}_2 \in \hat{\mathcal{Y}}_2$ and $v_2 \in \mathcal{V}_2$ we define

$$p(\hat{y}_2^n|x_2^n, v_2^n) = \sum_{x_0, y_0, y_1, y_2} p(x_0|x_1, x_2, v_1, v_2) p(y_0, y_1, y_2|x_0, x_1, x_2) p(\hat{y}_2|x_2, v_2, y_2)$$

and label them as $\hat{y}_2^n(z_2|w_{22}, w_{02})$, where $z_2 \in [1, 2^{n\hat{R}_2}]$.
***Partitioning:***

1. Create a partition over the set $\{1,2,\dots,2^{nR_{21}}\}$ with $2^{nR_{011}}$ disjoint cells $\mathcal{S}_{w_{011}}$ for $w_{011} \in \{1,2,\dots,2^{nR_{011}}\}$, each containing $2^{n(R_{21}-R_{011})}$ elements, and a partition over the set $\{1,2,\dots,2^{nR_{22}}\}$ with $2^{nR_{021}}$ disjoint cells $\mathcal{S}_{w_{021}}$ for $w_{021} \in \{1,2,\dots,2^{nR_{021}}\}$, each containing $2^{n(R_{22}-R_{021})}$ elements.
2. Randomly partition the set $\{1,2,\dots,2^{n\hat{R}_1}\}$ into $2^{nR_{012}}$ cells $\mathcal{S}_{w_{012}}$ for $w_{012} \in \{1,2,\dots,2^{nR_{012}}\}$, and the set $\{1,2,\dots,2^{n\hat{R}_2}\}$ into $2^{nR_{022}}$ cells $\mathcal{S}_{w_{022}}$ for $w_{022} \in \{1,2,\dots,2^{nR_{022}}\}$.

*Remark 5:* The first partition is referred to as deterministic partition and we use it for the restricted decoding [13].

In the joint decoding procedure, the decoder decodes a pair of bin numbers for partition 1, which their contents are jointly $\epsilon$-typical.

***Encoding:***

Let $w^{(b)}$ be the new message to be sent in block $b$. Assume that $\left(\hat{y}_1^n(z_1^{(b-1)}|w_{21}^{(b-1)}, w_{01}^{(b-1)}), y_1^{n(b-1)}, x_1^n(w_{01}^{(b-1)}), v_1^n(w_{21}^{(b-1)}, w_{01}^{(b-1)})\right)$ are jointly $\epsilon$-typical, and $\left(\hat{y}_2^n(z_2^{(b-1)}|w_{02}^{(b-1)}), y_2^{n(b-1)}, x_2^n(w_{02}^{(b-1)}), v_2^n(w_{22}^{(b-1)}, w_{02}^{(b-1)})\right)$ are jointly $\epsilon$-typical. Then, the codewords transmitted by the first and the second relays in block $b$ are

$$x_1^n(\widehat{w}_{01R_1}^{(b)}) = x_1^n\left(w_{011}(\widehat{w}_{21R_1}^{(b-1)}), w_{012}(z_1^{(b-1)})\right)$$

and

$$x_2^n(\widehat{w}_{02R_2}^{(b)}) = x_2^n\left(w_{021}(\widehat{w}_{22R_2}^{(b-1)}), w_{022}(z_2^{(b-1)})\right),$$

respectively, while the codeword transmitted by the sender is

$$x_0^n(w_1^{(b)}, w_{21}^{(b)}, w_{22}^{(b)}, w_{01}^{(b)}, w_{02}^{(b)})$$
$$= x_0^n\left(w_1^{(b)}, w_{21}^{(b)}, w_{22}^{(b)}, \left(w_{011}(w_{21}^{(b-1)}), w_{012}(z_1^{(b-1)})\right), \left(w_{021}(w_{22}^{(b-1)}), w_{022}(z_2^{(b-1)})\right)\right)$$

***Decoding:***

*At the first relay:* At the end of block $b$, the first relay estimates $\widehat{w}_{21R_1}^{(b)}$ such that $\left(x_1^n(\widehat{w}_{01R_1}^{(b)}), v_1^n(\widehat{w}_{21R_1}^{(b)}, \widehat{w}_{01R_1}^{(b)}), y_1^{n(b)}\right)$ are jointly $\epsilon$-typical. Using packing lemma [14], for sufficiently large $n$, the decoding error in this step is arbitrarily small if

$$R_{21} < I(V_1; Y_1|X_1) \quad (20)$$

Thus, the sender and the first relay know $w_{21}^{(b)}$ and they cooperate in sending $w_{011}^{(b+1)} = w_{011}(\widehat{w}_{21R_1}^{(b)})$ and $w_{011}^{(b+1)} = w_{011}(w_{21}^{(b)})$ by the first relay and the sender at the block $b+1$, respectively.

The first relay upon receiving $y_1^{n(b)}$, decides that $z_1^{(b)}$ is received if $\left(\hat{y}_1^n(z_1^{(b)}|\widehat{w}_{21R_1}^{(b)}, \widehat{w}_{01R_1}^{(b)}), y_1^{n(b)}, x_1^n(\widehat{w}_{01R_1}^{(b)}), v_1^n(\widehat{w}_{21R_1}^{(b)}, \widehat{w}_{01R_1}^{(b)})\right)$ are jointly $\epsilon$-typical. Using covering lemma [14], for sufficiently large $n$, the decoding error in this step is arbitrarily small if

$$\hat{R}_1 > I(\hat{Y}_1; Y_1|X_1, V_1) \quad (21)$$

*At the second relay:* At the end of block $b$, the second relay estimates $\widehat{w}_{22R_2}^{(b)}$ such that $\left(x_2^n(\widehat{w}_{02R_2}^{(b)}), v_2^n(\widehat{w}_{22R_2}^{(b)}, \widehat{w}_{02R_2}^{(b)}), y_2^{n(b)}\right)$ are jointly $\epsilon$-typical. Using packing lemma [14], for sufficiently large $n$, the decoding error in this step is arbitrarily small if

$$R_{22} < I(V_2; Y_2|X_2) \quad (22)$$

Thus, the sender and the second relay know $w_{22}^{(b)}$ and they cooperate in sending $w_{021}^{(b+1)} = w_{021}(\widehat{w}_{22R_2}^{(b)})$ and $w_{021}^{(b+1)} = w_{021}(w_{22}^{(b)})$ by the second relay and the sender at the block $b+1$, respectively.

The second relay upon receiving $y_2^{n(b)}$, decides that $z_2^{(b)}$ is received if $\left(\hat{y}_2^n(z_2^{(b)}|\widehat{w}_{22R_2}^{(b)}, \widehat{w}_{02R_2}^{(b)}), y_2^{n(b)}, x_2^n(\widehat{w}_{02R_2}^{(b)}), v_2^n(\widehat{w}_{22R_2}^{(b)}, \widehat{w}_{02R_2}^{(b)})\right)$ are jointly $\epsilon$-typical. Using covering lemma [14], for sufficiently large $n$, the decoding error in this step is arbitrarily small if

$$\hat{R}_2 > I(\hat{Y}_2; Y_2|X_2, V_2) \quad (23)$$

*At the receiver:* At the end of block $b+1$, the receiver looks for a unique pair $\left(\widehat{w}_{01D}^{(b+1)}, \widehat{w}_{02D}^{(b+1)}\right)$ such that $\left(x_1^n(\widehat{w}_{01D}^{(b+1)}), x_2^n(\widehat{w}_{02D}^{(b+1)}), y_0^{n(b+1)}\right)$ are jointly $\epsilon$-typical. Using packing lemma [14] and joint decoding [14, 11], for sufficiently large $n$, the decoding error in this step is arbitrarily small if

$$R_{011} + R_{012} < I(X_1; Y_0, X_2) \quad (24)$$
$$R_{021} + R_{022} < I(X_2; Y_0, X_1) \quad (25)$$
$$R_{011} + R_{012} + R_{021} + R_{022} < I(X_1, X_2; Y_0) \quad (26)$$

Then, the receiver considers block $b$ and estimates $\widehat{w}_{21D}^{(b)}$ and $\widehat{w}_{21D}^{(b)}$ such that $\left(x_1^n(\widehat{w}_{01D}^{(b+)}), v_1^n(\widehat{w}_{21D}^{(b)}, \widehat{w}_{01D}^{(b)}), y_0^{n(b)}\right)$ are jointly $\epsilon$-typical, and $\left(x_2^n(\widehat{w}_{02D}^{(b+)}), v_2^n(\widehat{w}_{22D}^{(b)}, \widehat{w}_{02D}^{(b)}), y_0^{n(b)}\right)$ are jointly $\epsilon$-typical. For sufficiently large $n$, the decoding error in this step is arbitrarily small if

$$R_{21} < I(V_1; Y_0|X_1) + R_{011} \quad (27)$$
$$R_{22} < I(V_2; Y_0|X_2) + R_{021} \quad (28)$$

Here, $\widehat{w}_{01D}^{(b)}$ and $\widehat{w}_{02D}^{(b)}$ were already determined in decoding steps similar to those for (24)-(26), and the decoding is restricted to $\widehat{w}_{21D}^{(b)}$ and $\widehat{w}_{22D}^{(b)}$ inside cells $\widehat{w}_{011}^{(b+1)}$ and $\widehat{w}_{021}^{(b+1)}$, respectively. This step is similar to the restricted decoding in [13].

Then, the receiver considers block $b$ again and calculates his ambiguity sets $\mathcal{L}_{1D}(y_0^{n(b)})$ and $\mathcal{L}_{2D}(y_0^{n(b)})$ of $\hat{z}_{1D}^{(b)}$ and $\hat{z}_{2D}^{(b)}$ such that $\hat{z}_{1D}^{(b)} \in \mathcal{L}_{1D}(y_0^{n(b)})$ and $\hat{z}_{2D}^{(b)} \in \mathcal{L}_{2D}(y_0^{n(b)})$ if $\left(x_1^n(\widehat{w}_{01D}^{(b)}), v_1^n(\widehat{w}_{21D}^{(b)}, \widehat{w}_{01D}^{(b)}), y_0^{n(b)}, \hat{y}_1^n\left(z_{1D}^{(b)}|\widehat{w}_{21D}^{(b)}, \widehat{w}_{01D}^{(b)}\right)\right)$

are jointly $\epsilon$-typical, and $\left(x_2^n(\hat{w}_{02D}^{(b)}), v_2^n(\hat{w}_{22D}^{(b)}, \hat{w}_{02D}^{(b)}), \hat{y}_2^n(z_{2D}^{(b)}|\hat{w}_{22D}^{(b)}, \hat{w}_{02D}^{(b)}), y_0^{n(b)}\right)$ are jointly $\epsilon$-typical. The receiver declares that $\hat{z}_{1D}^{(b)}$ and $\hat{z}_{2D}^{(b)}$ were sent in block $b$ iff there are unique

$$\hat{z}_{1D}^{(b)} \in \mathcal{S}_{\hat{w}_{012}^{(b)}} \cap \mathcal{L}_{1D}(y_0^{n(b)})$$

and

$$\hat{z}_{2D}^{(b)} \in \mathcal{S}_{\hat{w}_{022}^{(b)}} \cap \mathcal{L}_{2D}(y_0^{n(b)}),$$

respectively. For sufficiently large $n$, the decoding error in this step is arbitrarily small if

$$\hat{R}_1 < I(\hat{Y}_1; Y_0|X_1, V_1) + R_{012} \quad (29)$$
$$\hat{R}_2 < I(\hat{Y}_2; Y_0|X_2, V_2) + R_{022} \quad (30)$$

The proofs of (29) and (30) are similar to those in [1] for the compressed information of the relay. For a similar proof, refer to [11].

Then, the receiver declares that $\hat{w}_{1D}^{(b)}$ was sent in block $b$ if $\left(x_1^n(\hat{w}_{01D}^{(b)}), x_2^n(\hat{w}_{02D}^{(b)}), v_1^n(\hat{w}_{21D}^{(b)}, \hat{w}_{01D}^{(b)}), v_2^n(\hat{w}_{22D}^{(b)}, \hat{w}_{02D}^{(b)}), x_0^n(\hat{w}_{1D}^{(b)}, \hat{w}_{21D}^{(b)}, \hat{w}_{22D}^{(b)}, \hat{w}_{01D}^{(b)}, \hat{w}_{02D}^{(b)}), \hat{y}_1^n(z_{1D}^{(b)}|\hat{w}_{21D}^{(b)}, \hat{w}_{01D}^{(b)}), y_0^{n(b)}, \hat{y}_2^n(z_{2D}^{(b)}|\hat{w}_{22D}^{(b)}, \hat{w}_{02D}^{(b)})\right)$ are jointly $\epsilon$-typical. Using packing lemma [14], for sufficiently large $n$, the decoding error in this step is arbitrarily small if

$$R_1 < I(X_0; Y_0, \hat{Y}_1, \hat{Y}_2|X_1, X_2, V_1, V_2) + I(X_1, V_1; X_2, V_2) \quad (31)$$

*At the sender:* The sender upon receiving $y_1^{n(b)}$ and $y_2^{n(b)}$ via the feedback links, decides that the pair $(z_1^{(b)}, z_2^{(b)})$ is received if $\left(x_1^n(w_{01}^{(b)}), x_2^n(w_{02}^{(b)}), v_1^n(w_{21}^{(b)}, w_{01}^{(b)}), v_2^n(w_{22}^{(b)}, w_{02}^{(b)}), y_1^{n(b)}, y_2^{n(b)}, \hat{y}_1^n(z_1^{(b)}|w_{21}^{(b)}, w_{01}^{(b)}), \hat{y}_2^n(z_2^{(b)}|w_{22}^{(b)}, w_{02}^{(b)})\right)$ are jointly $\epsilon$-typical. Using multivariate covering lemma [14], for sufficiently large $n$, the decoding error in this step is arbitrarily small if

$$\hat{R}_1 > I(\hat{Y}_1; Y_1, Y_2, X_2, V_2|X_1, V_1) \quad (32)$$
$$\hat{R}_2 > I(\hat{Y}_2; Y_1, Y_2, X_1, V_1|X_2, V_2) \quad (33)$$
$$\hat{R}_1 + \hat{R}_2 >$$
$$I(\hat{Y}_1; Y_1, Y_2, X_2, V_2|X_1, V_1) + I(\hat{Y}_2; \hat{Y}_1, Y_1, Y_2, X_1, V_1|X_2, V_2) \quad (34)$$

Combining (20)-(30), (32), and (33), and applying Fourier-Motzkin elimination or the algorithm introduced in [15] and eliminating redundant inequalities, constraints (6)-(8) are derived.

## V. CONCLUSION

In this paper, we considered a relay network with two relays and partial feedback configuration from the relays to the transmitter. We presented two achievable rates. The first rate was achieved base on using the compress-and-forward coding scheme and the second rate was based on the combination of the decode-and-forward and the compressed-and-forward coding schemes. We showed how the feedback links make cooperation possible between the sender and the relays via transmitting the compressed information to the receiver.

ACKNOWLEDGMENT

The authors wish to thank the ISSL of Sharif University of Technology and CCL of K.N. Toosi University of Technology members for their helpful comments.